\documentclass[11pt,a4paper]{article}
\usepackage{jcappub}

\usepackage{amsmath,amssymb,bm,float,mathrsfs,subfigure,tabularx}
\usepackage{graphicx}
\usepackage{dcolumn}
\usepackage{hyperref}

  \newcommand{\beq}{\begin{equation}}
  \newcommand{\eeq}{\end{equation}}
  \newcommand{\al}[1]{\begin{align} #1 \end{align}}
  \newcommand{\bi}{\begin{itemize}}
  \newcommand{\ei}{\end{itemize}}
  \def\dd{\mathrm{d}}
  
\def\tMp{\widetilde{M}_{\rm Pl}}
\def\MM{M_{*}}
\def\M{{\cal M}}

\title{Modified gravity inside astrophysical bodies}

\author[a]{Ryo Saito}
\author[b]{Daisuke Yamauchi}
\author[c]{Shuntaro Mizuno}
\author[d,e]{J\'er\^ome Gleyzes}
\author[a]{David Langlois}

\affiliation[a]{%
APC (CNRS-Universite Paris 7), 10 rue Alice Domon et L\'eonie Duquet, 75205 Paris, France
}%

\affiliation[b]{%
Research Center for the Early Universe, Graduate School of Science, 
The University of Tokyo, Bunkyo-ku, Tokyo 113-0033, Japan
}%

\affiliation[c]{%
Waseda Institute for Advanced Study, Waseda University, Tokyo 169-8050, Japan
}%

\affiliation[d]{%
 CEA, IPhT, 91191 Gif-sur-Yvette c\'edex, France \\ [0.05cm]
CNRS,  URA-2306, 91191 Gif-sur-Yvette c\'edex, France
}%

\affiliation[e]{%
 Universit\'e Paris Sud, 15 rue George Cl\'emenceau, 91405,  Orsay, France
}%

\emailAdd{rsaito at apc.univ-paris7.fr}
\emailAdd{yamauchi at resceu.s.u-tokyo.ac.jp}
\emailAdd{shuntaro.mizuno at aoni.waseda.jp}
\emailAdd{jerome.gleyzes at cea.fr}
\emailAdd{langlois at apc.univ-paris7.fr}

\abstract{Many theories of modified gravity, including the well studied Horndeski models, are characterized by a screening mechanism that ensures that standard gravity is recovered near astrophysical bodies. 
In a recently introduced class of gravitational theories that goes beyond Horndeski, it has been found that new derivative interactions lead to a partial breaking of the Vainshtein screening mechanism
{\it inside} any gravitational source, although not outside.
We study the impact of this new type of deviation from standard gravity  on the density profile of a spherically symmetric matter distribution, in the nonrelativistic limit. For simplicity, we consider a polytropic equation of state and derive the modifications to the standard Lane-Emden equations. 
We also show the existence of  a universal upper bound on the amplitude of this type of modified gravity, independently of the details of the equation of state.
}

\begin{document}

\maketitle

\section{Introduction}
\label{sec:Introduction}

It is one of the biggest challenges of modern cosmology to understand the physical origin of the cosmic acceleration of the Universe discovered in Refs.~\cite{Riess:1998cb,Perlmutter:1998np}. It might eventually require the presence of a new type of matter, usually called dark energy, or alternatively a modification of
general relativity on cosmological scales. In addition to the standard tensor modes, models of the latter type often involve a new scalar degree of freedom that accounts for  cosmic acceleration. 
Among them, the simplest modification is described by a scalar-tensor theory where a single scalar degree of freedom is added (see e.g. \cite{FM}).
A very general class of scalar-tensor actions was found in \cite{Horndeski:1974wa} under the assumption that it leads to second-order equations of motion for both the metric and the scalar field.
These theories,  usually called Horndeski theories, include a large number of concrete models of modified gravity as specific cases.
Since  modifications of gravity are strongly constrained by precision tests of gravity on small scales, 
such as in the Solar system, any modified gravity model involving an additional propagating degree of freedom must include a mechanism that suppresses any potential fifth force that might appear on small scales.
The Vainshtein mechanism~\cite{Vainshtein:1972sx} represents one of the screening mechanisms, which is known to operate in models with nonlinear derivative interactions for the scalar mode.
These interactions can be large in the vicinity of a gravitational source, leading to self-screening of the fifth force.
The general theory exhibiting Vainshtein screening mechanism for the Horndeski Lagrangian has been
studied in the literature~\cite{Narikawa:2013pjr,Koyama:2013paa,Kimura:2011dc}
(see also \cite{DeFelice:2011th,Kase:2013uja}).

Until recently, Horndeski theories were believed to be the most general healthy scalar-tensor theories because of the generic occurence of Ostrogradski instabilities in systems with higher order equations of motion. 
However, it was proposed in Refs.~\cite{Gleyzes:2014dya, Gleyzes:2014qga} that Horndeski theories could be further  generalized  without introducing dangerous instabilities,  after a  detailed study of the true propagating degrees of freedom (see also \cite{Lin:2014jga} for a confirmation of these results). 
As found in Ref.~\cite{Kobayashi:2014ida}, a specific  feature of this new class of theories
is that it exhibits a deviation from general relativity inside a region filled with matter, while the usual Vainshtein screening is recovered just outside. 

Considering a spherical object with a radial density profile  denoted $\rho(r)$, the radial gravitational force in the nonrelativistic regime can be expressed as
\al{
	\frac{\dd\Phi}{\dd r} 
		=G_{\rm N}\left( \frac{\M}{r^2}-\epsilon\frac{\dd^2 \M}{\dd r^2} \right)
	\,,\label{eq:modified Poisson equation}
}
where $\M(r)=4\pi\int_0^r {r'}^2\rho (r')\dd r'$ is the mass enclosed in the sphere of radius $r$. The last term on the right hand side represents the deviation with respect to Newton's law. It is proportional to the radial derivative of the density, i.e. $\dd\rho/\dd r$ and its amplitude is characterized by the parameter 
$\epsilon$, which can be  written explicitly in terms of the functions that appear in the Lagrangian beyond Horndeski (See Ref.~\cite{Kobayashi:2014ida} and 
Appendix \ref{sec:Expression of epsilon parameter} for details). Although our main motivation for studying the gravitational law (\ref{eq:modified Poisson equation}) comes from the models beyond Horndeski introduced in Refs.~\cite{Gleyzes:2014dya, Gleyzes:2014qga}, the main discussion in this paper is independent of the underlying gravitational theory and thus applies to any theory leading to a phenomenological modification of this type.

The expected consequence of a deviation of Newton's law is a modification of the internal structure of any astrophysical body. This implies that modified gravity can in principle be constrained from observations of astrophysical objects whose internal physics is  well understood, such as stars.
The goal of the present work is to study, without entering into the astrophysical details, how the density  profile of a spherical object would be qualitatively modified by a  gravitational force of the form  (\ref{eq:modified Poisson equation}). 

For simplicity, we will assume the equation of state to be polytropic. 
In standard gravity, the density profile for polytropic equations of state is determined by solutions of the so-called Lane-Emden equation \cite{1939isss.book.....C}. With the new gravitational interaction given in (\ref{eq:modified Poisson equation}), we obtain a {\it modified} Lane-Emden equation, which can be solved numerically. We also show that, independently of the equation of state for matter, there is a universal upper bound on the value of $\epsilon$ in order to obtain physically sensible solutions. 

This paper is organized as follows: In section \ref{sec:Structure of a star in beyond Horndeski theory}, 
we derive  the {\it modified} Lane-Emden equation. We then show the existence of a critical value for $\epsilon$.
In the subsequent section, we  perform the numerical integration of 
the modified Lane-Emden equation. We also construct an exact analytic 
solution for a specific value of the polytropic index.
Section \ref{sec:Summary} is devoted to our conclusions.

\section{Density profile in modified gravity}
\label{sec:Structure of a star in beyond Horndeski theory}

\subsection{ Modified Lane-Emden equation}

In this section, we consider a static, spherically symmetric distribution of matter as
a simple toy model for an astrophysical object such as a non-relativistic star.
Although this model is too simple to be directly confronted  with observational data, it enables us to capture the essential modifications of the stellar structure and to provide simple estimates of the novel effects due to the breaking of the Vainshtein screening.
For a static, spherically symmetric non-relativistic source, the hydrostatic  equation reads
\al{
	\frac{\dd P}{\dd r}
		=-\rho\frac{\dd\Phi}{\dd r}
		=-G_{\rm N}\rho\left(\frac{\M}{r^2}-\epsilon\frac{\dd^2 \M}{\dd r^2}\right)
	\,,\label{eq:Euler eq}
}
where $P$ is the pressure and we have used the modified gravitational equation \eqref{eq:modified Poisson equation}.
Except the latter,  all other equations governing the matter distribution are supposed to be unchanged.
Note that in our treatment,  $\epsilon$ is {\it a priori} assumed to take any value,   both positive or negative.
(see Appendix \ref{sec:Expression of epsilon parameter} for the possible values of $\epsilon$ in the context of the theories beyond Horndeski.)

In order to get rid of the integral function $\M(r)$, one can multiply (\ref{eq:Euler eq}) by $r^2/\rho$ and then  take its derivative, which yields
\al{
	\frac{\dd}{\dd r}\left(\frac{r^2}{\rho}\frac{\dd P}{\dd r}\right)
		+4\pi G_{\rm N} r^2\left[\left(1-6\epsilon\right)\rho  -6\,\epsilon\, r \frac{\dd\rho}{\dd r} -\epsilon \, r^2\frac{\dd^2\rho}{\dd r^2}\right]=0
	\,.\label{eq:eq2}
}
To obtain a closed equation and be able to solve it explicitly for the radial profile, one needs to specify the equation of state for matter.
Here,  we simply assume a polytropic equation of state of the form 
\al{
	P = K \rho^{1+\frac{1}{n}}
	\,,\label{eq:pol}
}
where $K$ and $n$ are positive constants.

Following the standard procedure, we introduce the following dimensionless variable $\xi$ and function $\chi$ such that
\al{
	\xi =\frac{r}{r_{\rm c}}
	\,,\ \ \ 
	\rho =\rho_{\rm c}\, [\chi (\xi )]^n
	\,,
	\label{eq:rho_chi}
}
where $\rho_{\rm c}$ denotes the energy density at the center and
\al{
	r_{\rm c}=\sqrt{\frac{(n+1)K\rho_c^{-1+\frac{1}{n}}}{4\pi G_{\rm N}}}
	\,.\label{eq:r_c def}
}
Using these quantities and combining eqs.~\eqref{eq:Euler eq}--\eqref{eq:r_c def}, after straightforward calculations, 
we obtain a {\it modified} Lane-Emden equation, given by
\al{
	\frac{1}{\xi^2}\frac{\dd}{\dd\xi}\biggl[\xi^2\frac{\dd}{\dd\xi}\Bigl(\chi -\epsilon\xi^2\chi^n\Bigr)\biggr]
		=-\chi^n
	\,.\label{eq:mLEeq}
}
For $\epsilon =0$\,, one recognizes the standard Lane-Emden equation.

The density profile can be obtained by solving this equation with the following boundary conditions for $\chi (\xi)$ at the center of a star $\xi =0$ : $\chi (0)=1$, $\left(\dd\chi /\dd\xi\right)\bigl|_{\xi =0}=0$\,.
Once the solution of this equation is computed, the radius and mass of the object, respectively denoted $R$ and $M$, are 
determined by the first zero $\xi_1$ of the function $\chi$, i.e. $\chi(\xi_1) = 0$, through
$R=r_{\rm c}\xi_1$ and the integration of the density up to the radius, 
\al{
	M=4\pi r_{\rm c}^3\rho_{\rm c}\int_0^{\xi_1}\xi^2[\chi (\xi )]^n\dd\xi
	\,.
\label{def_mass}}
Note that $R/r_{\rm c}$ and $M/4\pi r_{\rm c}^3\rho_{\rm c}$ are uniquely determined for a given value of $\epsilon$.
Therefore, the usual scaling $M \propto R^{\frac{3-n}{1-n}}$ is unchanged, although the proportionality coefficient is modified, as it depends on $\epsilon$.
Using eq.~\eqref{eq:mLEeq} and the definition of $\xi_1$, 
the integration for the mass (\ref{def_mass}) can be analytically performed and gives
\al{
	M&=
	\begin{cases}
	\displaystyle{-4\pi r_{\rm c}^3\rho_{\rm c}\xi_1^2\frac{\dd\chi}{\dd\xi}\biggl|_{\xi =\xi_1}\, , \quad n>1}\\
	\displaystyle{-4\pi r_{\rm c}^3\rho_{\rm c}\xi_1^2\frac{\dd\chi}{\dd\xi}\biggl|_{\xi =\xi_1}\Bigl( 1 - \, \epsilon \xi_1^2  \Bigr)\, , \quad n=1}
	\end{cases}
	\,.
\label{analytic_mass}
}
For $n>1$,  this result is independent of $\epsilon$ and coincides with the  expression obtained from the standard Lane-Emden equation.
By contrast, the mass diverges for $n<1$, which indicates  that the modification of gravity leads to a dramatic change in this case. In the following, we will  assume $n \ge 1$.

\subsection{ Universal bound on $\epsilon$}

As we will see explicitly in the next section by solving numerically the modified Lane-Emden equation, it turns out that there exists a critical value for the parameter $\epsilon$ beyond which one cannot find a meaningful profile. As we now show, this critical value for $\epsilon$ can be derived from very general arguments that are independent of the details of the equation of state. 

Any nonsingular solution is characterized by  density and pressure profiles with finite values and vanishing first derivatives at the center. This means that one can expand $\rho(r)$ and $P(r)$ near the center $r=0$, as
\beq
\rho=\rho_{\rm c} +\frac12 \rho_2 \frac{r^2}{R^2}+\cdots\,, \qquad P=P_{\rm c}+\frac12 P_2 \frac{r^2}{R^2}+\cdots\,.
\eeq
Inserting these expansions into (\ref{eq:eq2}), one obtains, from the term at lowest order in $r$, the relation
\beq
\label{eq:Ppositive}
P_2=-\frac{4\pi G_{\rm N} \rho_{\rm c}^2 R^2}{3}(1-6\epsilon)\,.
\eeq
For any physically reasonable equation of state, one expects the density and pressure to decrease when going away from the center, which means in particular $P_2<0$. The above relation shows that this is possible only if $\epsilon<1/6$. 

Physically, this can be understood by noting that gravity becomes effectively repulsive at the center of the object when $1-6\epsilon <0$. Indeed, near the center, we have 
\beq
\M \simeq \frac{4\pi}{3} \rho_{\rm c}r^3\,, \qquad \frac{\dd^2 \M}{\dd r^2}  \simeq 6\frac{\M}{r^2}\,.
\eeq
Substituting these into the right hand side of (\ref{eq:modified Poisson equation}), one finds that  the gravitational force is proportional to $(1-6\epsilon)$ and thus changes sign when $\epsilon$ crosses the critical value $1/6$. 

Given a physically reasonable equation of state for which the pressure increases with the density, the above result that the pressure increases with $r$ when $\epsilon>1/6$ implies that the density also increases with $r$ near the center. In fact, one can also show that  the density gradient cannot vanish at any higher radius (see Appendix \ref{a:ie} for details). Consequently,  this puzzling behavior is not confined to the core and continues for higher radii.  
We thus conclude that one cannot construct a physically sensible profile when $\epsilon >1/6$.

\section{ Solutions for the density profile}

\subsection{ Numerical solutions}
Let us first discuss the asymptotic behavior of $\chi$ around $\xi =0$ 
to avoid the numerical instability at $\xi =0$.
Expanding $\chi$ around the center of the core with the appropriate boundary condition,
namely $\chi (\xi )=\sum_{m=0}c_m\xi^m$ with $c_0=1\,,c_1=0$\,, 
we find that the following coefficients solve the equation \eqref{eq:mLEeq},
\al{
	c_2=-\frac{1}{6}+\epsilon
	\,,\ \ 
	c_3=0
	\,,\ \ 
	c_4=\frac{n}{120}\left( 1-6\epsilon\right)\left( 1-20\epsilon\right)
	\,,\ \ 
	c_5=0
	\,,\cdots
	\,.
\label{asymp_center}}
We note that the nonvanishing coefficients of $c_m$ approach to zero as $\epsilon \to 1/6$. 
Hence we should take care of the convergence when performing the numerical integration.

We now present the results obtained by solving the modified Lane-Emden equation (\ref{eq:mLEeq}) numerically.
We consider the polytropes with the index $n=3$ and $n=1$, 
which give models for main sequence stars and neutron stars, respectively. 
For these values of $n$, we investigate how the new gravitational force changes the density profile of a star for a given value of $\epsilon$.

	\begin{figure}[h!]
		\centering
		\begin{minipage}{.45\linewidth}
		\centering
		\includegraphics[width=\linewidth]{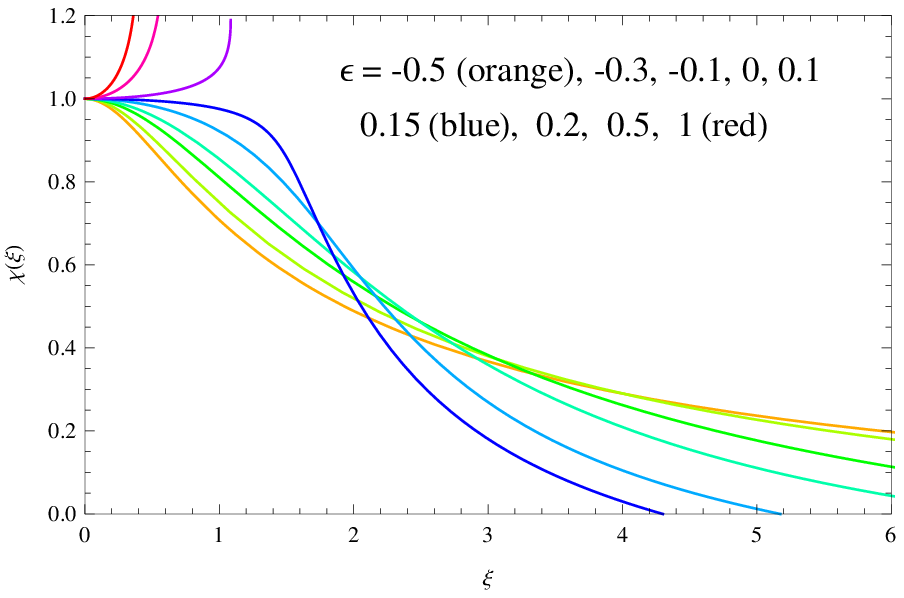}
		\end{minipage}
		\begin{minipage}{.45\linewidth}
		\centering
		\includegraphics[width=\linewidth]{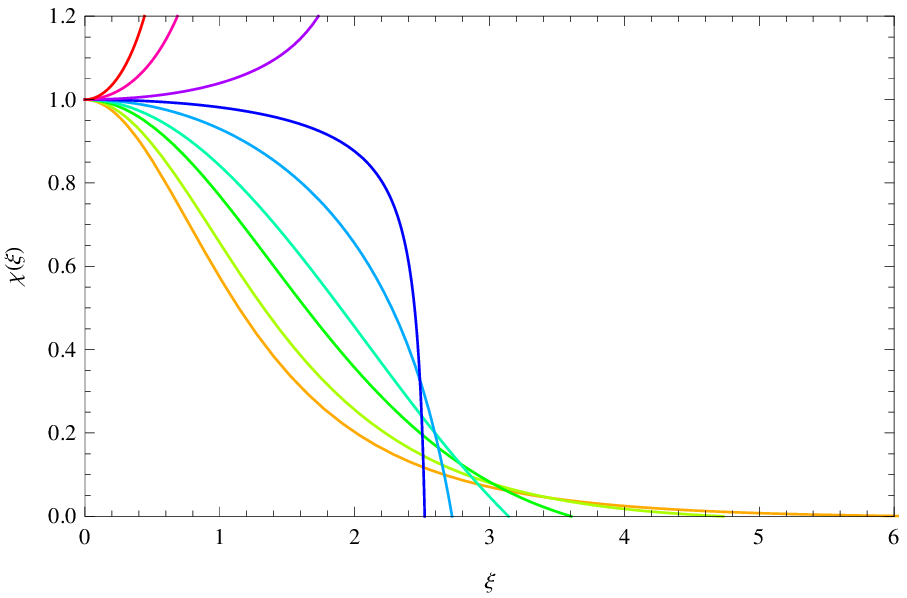}
		\end{minipage}
		\caption{Numerical results for the profile $\chi (\xi)$ with $n=3$ ({\it Left}) 
and the one with $n=1$  ({\it Right}).  Different color of the lines corresponds to the different value of $\epsilon$  as is
shown in the left pannel.}
		\label{fig:polytropic index}
	\end{figure}

\begin{figure}[h!]
			\centering
		\begin{minipage}{.40\linewidth}
		\centering
		\includegraphics[width=\linewidth]{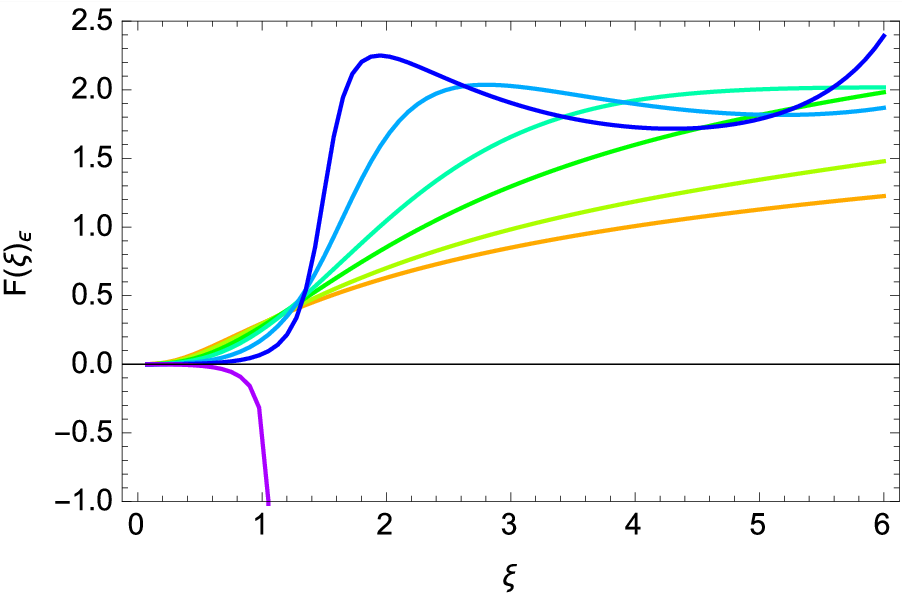}
		\end{minipage}
		\begin{minipage}{.40\linewidth}
		\centering
		\includegraphics[width=\linewidth]{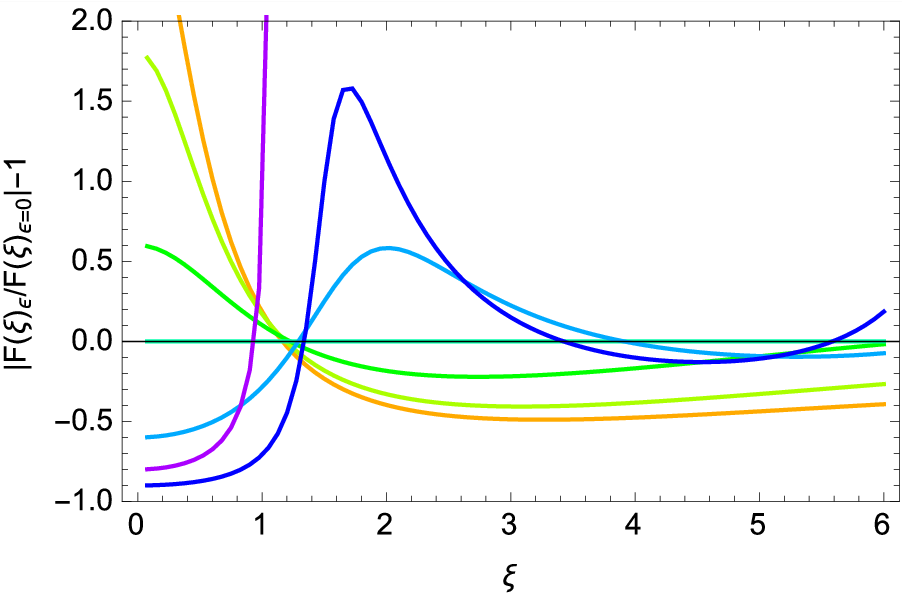}
		\end{minipage}
		\caption{The gravitational force and its sign ({\it Left}) and relative difference in the magnitude of the force ({\it Right}) for $n=3$ and various values of $\epsilon$, with the same colors as in Fig.~\ref{fig:polytropic index}. In particular, $\epsilon=0$ is the light blue line. Note that because of eqs.~\eqref{eq:Euler eq}, \eqref{eq:pol} and \eqref{eq:rho_chi}, the magnitude of the force is directly related to the steepness of the profiles in Fig.~\ref{fig:polytropic index}.}
\label{plot_Force}
	\end{figure}

In Fig.~\ref{fig:polytropic index}, we plot the solutions of the modified Lane-Emden equation 
with the polytropic index $n=3$ (Left)  and $n=1$ (Right) for several values of $\epsilon$.
We can find two features from these figures:
\begin{itemize}
\item For $\epsilon$ larger than the critical value $1/6$, the density never approaches to zero.
This implies that a physically sensible profile cannot be obtained, as expected from the discussion in the previous section where it was found that the force is repulsive in this case.
\item By contrast, for $\epsilon$ below the critical value $1/6$, the force is always attractive (see Fig. \ref{plot_Force}). One can see that 
the profile near the core tends to be steeper for smaller $\epsilon$, while it becomes flatter when moving further out. 
This indicates that gravity becomes stronger in the inner region but weaker in the outer region. 
\end{itemize}
Fig.~\ref{plot_Force} confirms that when $\epsilon$ is negative, the magnitude of the force
\beq
\label{eq:forcegrav}
|F_{\rm grav}|=G_{\rm N}\rho\left|\frac{\M}{r^2}-\epsilon\frac{\dd^2 \M}{\dd r^2}\right| \,,
\eeq
is larger than in standard gravity near the center, as expected since the boundary conditions imply $\dd^2\M/\dd r^2>0$. Further out, however, $\dd^2\M/\dd r^2$ becomes negative  as it  also contains $\dd\rho /\dd r$ which is negative. Consequently, the magnitude of the force is reduced with respect to the standard situation, thus leading to a larger radius of the object. In the case  $0<\epsilon<1/6$, one finds the opposite effects.

\begin{figure}[h!]
		\centering
		\begin{minipage}{.45\linewidth}
		\centering
		\includegraphics[width=\linewidth]{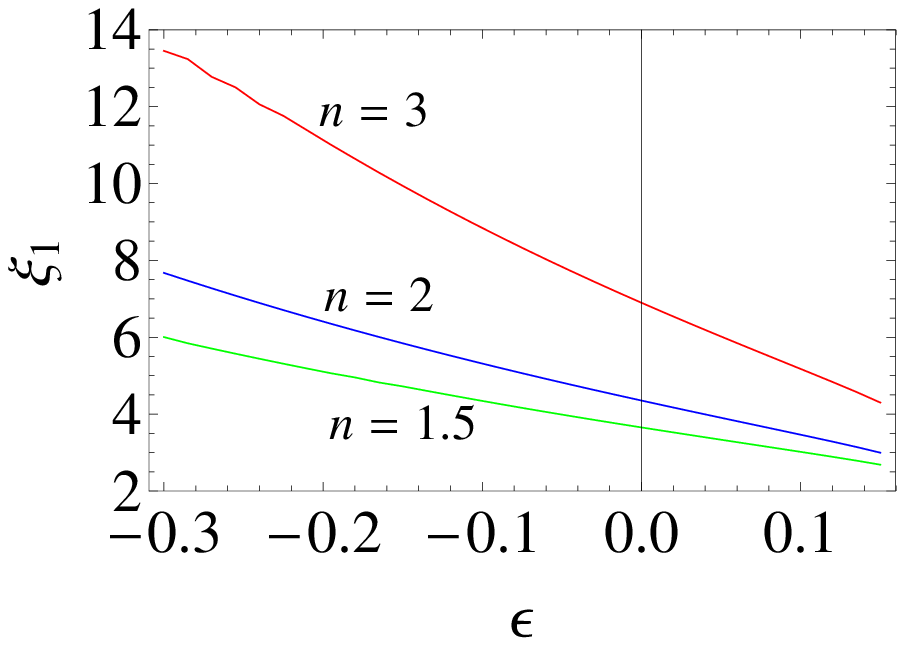}
		\end{minipage}
		\begin{minipage}{.45\linewidth}
		\centering
		\includegraphics[width=\linewidth]{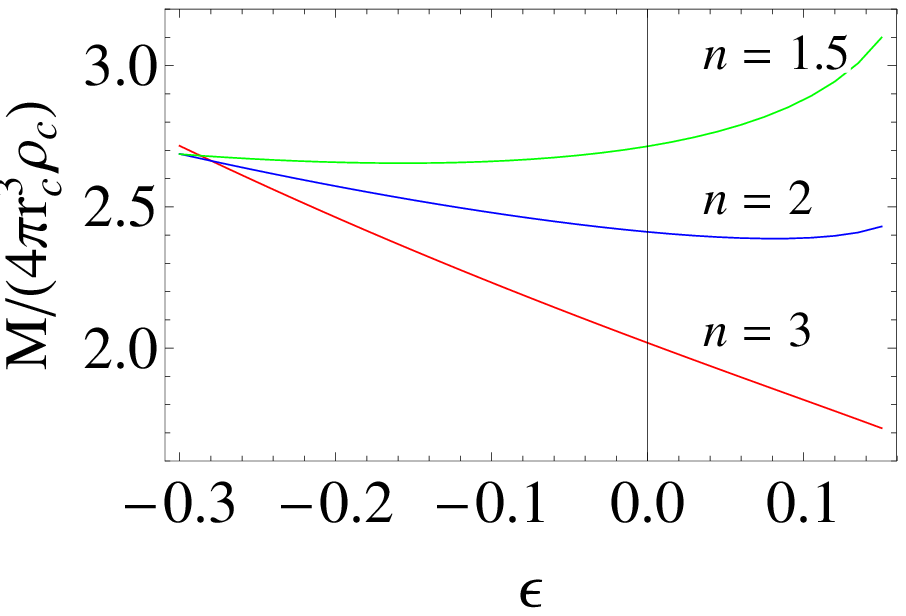}
		\end{minipage}
		\caption{Numerical results for the dimensionless radius $\xi_1=R/r_c$  ({\it Left})
and the dimensionless mass $M/(4\pi r_{\rm c}^3 \rho_{\rm c})$ ({\it Right}) as functions of $\epsilon$.  $\epsilon=0$ corresponds to the case with standard gravity.
As illustrative values of the polytropic index, we choose $n=1.5$ (green), $n=2$ (blue), and $n=3$ (red). }
\label{plot_epsilon_xi}
	\end{figure}

In Fig.~{\ref{plot_epsilon_xi}}, we plot $\xi_1=R/r_c$ and $M/(4\pi r_{\rm c}^3 \rho_{\rm c})$ as a function of $\epsilon$, for three equations of state ($n=1.5, 2$ and $3$. We observe that the dimensionless radius $R/r_c$ always decreases as $\epsilon$ increases. By contrast, the variation of the dimensionless mass $M/(4\pi r_{\rm c}^3 \rho_{\rm c})$ depends on the equation of state.

\subsection{ Analytic Solution}

In standard gravity, it is well known that the Lane-Emden equation can be solved analytically 
in the case $n=1$. The corresponding analytic solution is given by
\al{ \chi (\xi) = \frac{\sin \xi}{\xi}\,.
}
In this section, we show that analytic solutions can be found also for the modified Lane-Emden equation (\ref{eq:mLEeq}) in the case $n=1$.
The analytic solutions are helpful to point out the existence of the critical point more explicitly
as well as to check the numerical solutions obtained before.\

For $n=1$,  by changing the variable from $ \xi$ to $x$ defined by $x \equiv \epsilon \xi^2$, the equation (\ref{eq:mLEeq})
can be rewritten as
\begin{align}
x(1-x) \frac{{\rm d}^2 \chi }{{\rm d} x^2} + \left(\frac32 - \frac72 x  \right) \frac{{\rm d} \chi}{{\rm d} x} - \left(\frac{25}{16} - \frac{4+\epsilon}{16 \epsilon} \right) \chi =0\,.
\label{eq:gaussdiff}
	\end{align}
We  see that eq.~(\ref{eq:gaussdiff}) takes the form of Euler's hypergeometric differential equation
and the solution satisfying the boundary conditions is given by
\begin{align}
\chi (\xi)= {}_2 F_1 \left[ \frac54 - \frac14 \sqrt{\frac{4+ \epsilon}{\epsilon}},\;\;\frac54 + \frac14 \sqrt{\frac{4+ \epsilon}{\epsilon}},\;\;\frac32;\;\; \epsilon \xi^2  \right]\,,
\label{hypergeometric}
	\end{align}
where $ {}_2 F_1 (a,b,c;x)$ is the hypergeometric function.

As is expected, these solutions satisfy eq.~(\ref{asymp_center}) around the center.
Furthermore, we can see explicitly that  $ {}_2 F_1(0,5/2,3/2, \xi^2/6) = 1$ holds identically,
which confirms that the value $\epsilon=1/6$ corresponds to a threshold beyond which  one cannot find a physically sensible profile.
We have also checked that the profile $\chi(\xi)$ obtained numerically for $n=1$ coincides with the analytical expression (\ref{hypergeometric}).

\section{Summary} \label{sec:Summary} 
In this paper, we have investigated some consequences of a partial breaking of the Vainshtein screening mechanism inside an astrophysical object, which has been shown to arise in theories beyond Horndeski.
Assuming a polytropic equation of state, we have derived a diffential  equation for the density, which generalizes the Lane-Emden equation obtained in standard gravity [eq.~\eqref{eq:mLEeq}].
By solving this equation numerically, we have obtained the modified density profile, and, consequently,  the modified mass and radius of a spherical object. The modifications depend on the amplitude of the new gravitational effect, characterized by a parameter $\epsilon$.   We have found the existence of a critical value 
for $\epsilon$, below which viable profiles do not exist.
We have argued that this is also true for any physically reasonable equation of state, i.e. the equation of state where the density increases with pressure.
This result can be physically understood from the fact that gravity effectively becomes repulsive for $\epsilon > 1/6$.  This is why a stable distribution of matter cannot be reached.

In our analysis, we have considered a very simple set-up: a static, spherically symmetric and non-relativistic object with a polytropic equation of state. To make detailed comparisons with observations, and thus obtain precise constraints on the allowed amplitude of this new effect,  it would be necessary to refine our analysis by  including   a realistic equation of state, as well as rotation.
Relativistic corrections should also be taken into account  for compact stars.


\vspace{5ex}
\noindent
{\bf Note added}:
While this paper was in preparation, Ref.~\cite{Koyama:2015oma} appeared, which  also investigates the consequences of a partial breaking of the Vainshtein screening mechanism inside an astrophysical object. We note that they restricted their attention to a limited region of parameter space corresponding to $\epsilon <0$ in our notation.

\acknowledgments
We thank T.~Shigeyama and T.~Suda for valuable comments and useful suggestions.
R.S. is supported by JSPS Postdoctoral Fellowships for Research Abroad.
D.Y. is supported by Grant-in-Aid for JSPS Fellows (No.259800).
S.M. is supported by JSPS Grant-in-Aid for Research Activity Start-up
No. 26887042 and Waseda University Grant for Special Research Projects
 (Project number 2014S-191).


\appendix

\section{Expression of the $\epsilon$ parameter} \label{sec:Expression of epsilon parameter}

\subsection{Connection to the EFT parameters}

Let us be more explicit about the form of $\epsilon$ in eq.~\eqref{eq:modified Poisson equation}. For concreteness, we study the quartic Lagrangian
\beq
\begin{split}
L_4=G_4(\phi,X)R+G_{4X}(\phi,X)\left[(\Box\phi)^2-\phi_{\mu\nu}\phi^{\mu\nu}\right]-\frac{F_4(\phi,X)}2\epsilon^{\mu \nu \alpha\beta}\epsilon_{\mu' \nu' \alpha'\beta}\phi^{\mu'}\phi_\mu \phi_\nu^{\ \nu'}\phi_\alpha^{\ \alpha'} \, ,
\end{split}
\eeq
with the shorthand notation
\beq
\phi_\mu\equiv \nabla_\mu \phi\, , \quad \phi_{\mu\nu}\equiv \nabla_\nu\nabla_\mu \phi\,.
\eeq
The gradient of  the gravitational potential  is given in this case by \cite{Kobayashi:2014ida}
\beq
\frac{\dd\Phi}{\dd r} 
		=G_{\rm N}\left( \frac{M}{r^2}-\frac{\alpha_*^2}{4\pi \widetilde \M^2_{\rm Pl} G_{\rm N}\Xi}\frac{\dd^2 M}{\dd r^2} \right)\, ,
\eeq
where $\widetilde M_{\rm Pl}$ is a mass scale for the metic perturbations introduced in \cite{Kobayashi:2014ida}.
Comparing this and eq.~\eqref{eq:modified Poisson equation} gives the expression of $\epsilon$ in terms of
the model parameters:
\beq
\label{eq:espilonKWY}
\epsilon=\frac{\alpha_*^2}{4\pi \widetilde M_{\rm Pl}^2 G_{\rm N}\, \Xi}\,.
\eeq
The various terms are those defined in \cite{Kobayashi:2014ida} and take the specific form of
\begin{align}
&\frac{\tMp}{\Lambda^3} \alpha_* \equiv  XF_4\, , \quad (8\pi G_{\rm N})^{-1}\equiv 2G_4-8X(G_{4X}+XG_{4XX})-4X^2(5F_4+2XF_{4X})\, \\
&\Xi \equiv \mathcal{G}\left(4\alpha_1\alpha_2-2\alpha_1\alpha_*+ \mathcal{G}\nu\right)-2\mathcal{F}\alpha_1^2\, ,\quad \tMp^2\mathcal{G}\equiv2\left(G_4-2XG_{4X}\right)\, , \quad \tMp^2 \mathcal{F}\equiv2G_4
\end{align}
and
\begin{align}
\frac{\tMp}{\Lambda^3}\alpha_1&\equiv G_{4X}+2XG_{4XX}+X(5F_4+2XF_{4X}) \, ,\\
\frac{\tMp}{\Lambda^3}\alpha_2&\equiv G_{4X}+XF_4 \, ,\\
\frac{\nu}{\Lambda^6}&\equiv G_{4XX}+2F_4+XF_{4X} \, .
\end{align}
Here, we set the energy scale $\widetilde{M}_{\rm Pl}$ to be,
	\begin{align}
		\widetilde{M}_{\rm Pl} ^2=\MM^2&\equiv 2G_4-4XG_{4X}-4X^2F_4\, ,
	\end{align}
which corresponds to the mass scale $M$ in \cite{Gleyzes:2014qga}. 
It canonically normalizes the tensor perturbations and should be always positive to avoid ghost instability.

It is also convenient to reexpress $\epsilon$ in the language of the so-called EFT formalism, which has been applied to Horndeski theories in \cite{Gleyzes:2013ooa} and to their extensions in \cite{Gleyzes:2014qga} (see also \cite{Gleyzes:2014rba} for a recent review). In terms of the parameters used in \cite{Gleyzes:2014qga}, defined as
\begin{align}
\alpha_B&\equiv -\frac{4X}{\MM^2}\left[4XF_4+2X^2F_{4X}+G_{4X}+2XG_{4XX}\right]\, ,\\
\alpha_T&\equiv \frac{4X}{\MM^2}\left[G_{4X}+XF_4\right]\, ,\\
\alpha_H&\equiv \frac{4X^2}{\MM^2}F_4\,,
\end{align}
we find that the functions introduced earlier take the form
\begin{align}
\mathcal{F}= 1+\alpha_T&\, , \quad \mathcal{G}= 1+\alpha_H\, ,\quad \left(\frac{X}{M_{\ast}\Lambda^3}\right)\alpha_*=\frac{\alpha_H}4 \, , \\
G_{\rm N}&=\left[8\pi\MM^2(1+\alpha_B)\right]^{-1}\, ,
\end{align}
and
	\begin{align}
		& \left(\frac{X}{M_{\ast}\Lambda^3}\right)\alpha_1= \frac{\alpha_H-\alpha_B}{4}\, , \quad  \left(\frac{X}{M_{\ast}\Lambda^3}\right)\alpha_2= \frac{\alpha_T}4\, , \\
		& \left(\frac{X}{M_{\ast}\Lambda^3}\right)^2\nu=\frac{\alpha_H-\alpha_T-\alpha_B}8 \, .
	\end{align}
Finally, one can express eq.~\eqref{eq:espilonKWY} as
\beq
\epsilon=\frac{\alpha_H^2}{\alpha_H-\alpha_T-\alpha_B(1+\alpha_T)}\, .
\eeq

\subsection{Viable parameter region}
In the previous subsection, we expressed the $\epsilon$ parameter in terms of four EFT parameters. 
To avoid several inconsistencies, these parameters cannot be arbitrary. 
First, because the gravitational force around a star is attractive, 
the effective gravitational constant $G_{\rm N}$ should be positive,
	\begin{align}
		(8\pi G_{\rm N})^{-1} = M_{\ast}^2(1+\alpha_B)  > 0 \,,
	\end{align}
or 
	\begin{align}
		f_{\ast} \equiv (8\pi G_{\rm N} M_{\ast}^2)^{-1} = 1+\alpha_B > 0 \,. 
	\end{align}
Here, the dimensionless parameter $f_{\ast}$ can be understood as the strength of the graviton interaction with respect to that of the gravitational force around a star.
In addition, we should require that the speed of sound of the tensor perturbation should be positive,
	\begin{align}
		c_T^2 \equiv 1+ \alpha_T >0 \,,
	\end{align}
to avoid the gradient instability. In terms of these positive quantities, the $\epsilon$ parameter is expressed as,
	\begin{align}
		\epsilon=\frac{\alpha_H^2}{1+\alpha_H - f_{\ast}c_T^2}\, .
	\end{align}
Therefore, $\epsilon$ can be either positive or negative depending on the value of $\alpha_H$. \footnote{One must also ensure that there is no instabilities in the scalar sector. This leads to a rather intricate condition \cite{Gleyzes:2014qga}, which does not seem to restrict $\epsilon$ to be either positive or negative.} 

When $G_4 = M_{\rm Pl}^2/2$ and $F_4 = {\rm const}$, the $\epsilon$ parameter is simply expressed as,
 	\begin{align}
		\epsilon = \frac{X^2 F_4}{M_{\rm Pl}^2} ~.
	\end{align}
 In the $G^3$-galileon theory considered in \cite{Koyama:2015oma}, where $F_4 = -1/\Lambda^6$, it is always negative. The $\epsilon$ parameter corresponds to $4\Upsilon$ in \cite{Koyama:2015oma}. \footnote{There seems to be a typo in eq. (17) of \cite{Koyama:2015oma}. $\Upsilon/4$ will appear instead of $\Upsilon/8$ in eq. (23) and (24). }

\section{No physically sensible profile for $\epsilon > 1/6$}\label{a:ie}

For a physically reasonable equation of state where the density increases with pressure, eq.~\eqref{eq:Ppositive} implies that the density is also a monotonically increasing function of $r$ near the center when $\epsilon>1/6$. 
Here, we show that the density continues to increase away from the center and then no physically sensible profile is obtained.
Let us assume there exists a radius $r_*$ where the density turns to decrease, meaning $(\dd\rho /\dd r)(r_*)=0$. Since $\rho$ is assumed to have been strictly increasing for $r<r_{\ast}$, the following inequality holds
\beq
\label{eq:Mstar}
		\M (r_*)< \frac{4\pi}{3}\rho(r_*)r_*^3 \,,
\eeq
because $\rho(r)$ has the maximum at the outermost radius $r_*$. Moreover, by using the relation between mass and density, we also obtain,
	\begin{align}
		\frac{\dd^2\M}{\dd r^2} (r_*)= 4\pi\left[ 2\rho(r_*)r_* + \frac{\dd\rho}{\dd r}(r_*) r_*^2 \right] = 8\pi\rho(r_*) r_* \,.
	\end{align}
Inserting this, along with eq.~\eqref{eq:Mstar}, in the hydrostatic equation~\eqref{eq:Euler eq} yields
\beq
\frac{\dd P}{\dd r}\biggl|_{r =r_*}>-\frac{4\pi G_{\rm N}\, \rho^2 r_*}3(1-6\epsilon)>0 \,,
\eeq
which also implies $(\dd\rho /\dd r)(r_{\ast})>0$ under our assumption on the equation of state. This contradicts the assumption $(\dd\rho /\dd r)(r_*)=0$. Therefore, unless the relation between pressure and density is an exotic one, there is no possibility of having an astrophysical body when $\epsilon>1/6$.


\end{document}